\documentclass[conference]{IEEEtran}

\let\labelindent\relax

\usepackage[T1]{fontenc}
\usepackage[utf8]{inputenc}

\usepackage[british]{babel}
\usepackage[babel]{csquotes}
\MakeAutoQuote{“}{”}
\MakeAutoQuote*{‘}{’}

\usepackage{makeidx}  
\usepackage{times,amsmath,epsfig}
\usepackage{epstopdf}
\usepackage{amsmath}
\usepackage{lstsemantic}  
\usepackage{xcolor}
\usepackage{colortbl}
\usepackage{rotating}
\usepackage{enumerate}
\usepackage{url}
\usepackage{listings}
\usepackage{tabularx}
\usepackage[bookmarks=false]{hyperref}
\usepackage{pifont}

\usepackage{subfigure}

\usepackage{amssymb}
\usepackage{graphicx}
\usepackage{float}
\usepackage{algorithm}
\usepackage{multirow}

\usepackage{enumitem}
\setlist[itemize]{topsep=0pt,itemsep=-1ex,partopsep=1ex,parsep=1ex}

\newtheorem{Def1}{Definition}

\usepackage{listings}
\usepackage{color}
\definecolor{grey}{rgb}{0.9,0.9,0.9}
\lstnewenvironment{lqml}[2][]
  {
    \lstset{
      basicstyle=\scriptsize \ttfamily,
      breaklines=true,
      captionpos=b,
      frame=bottomline,
      extendedchars=true,
      caption=#1,
      label=#2,
      morekeywords={def,metric,label,description,match,action,map,count,typeof,finally,unique,actionadd,result,ratio},
      escapechar=@,
    }
  }
   {
  }

\usepackage[backend=bibtex,
firstinits=true,style=numeric,
urldate=iso8601,isbn=false,doi=false]{biblatex}
\DeclareNameAlias{author}{last-first}
\DeclareNameAlias{editor}{last-first}
\DeclareNameAlias{translator}{last-first}
\DeclareFieldFormat{labelnumberwidth}{#1.}
\DeclareFieldFormat{title}{#1}
\DeclareFieldFormat
  [article,inbook,incollection,inproceedings,patent,thesis,unpublished]
  {title}{#1}
\DeclareFieldFormat{journaltitle}{#1}
\newbibmacro*{volume+number+eid}{%
  \printfield{volume}%
  \iffieldundef{number}
    {}
    {(%
      \printfield{number}%
     )%
    }%
  \setunit{\addcomma\space}%
  \printfield{eid}}
\DeclareFieldFormat{url}{\url{#1}}

\begin{document}
\title{Luzzu -- A Framework for Linked Data Quality Assessment}

\author{
\IEEEauthorblockN{Jeremy Debattista, S\"{o}ren Auer, Christoph Lange}
\IEEEauthorblockA{Enterprise Information Systems, University of Bonn \& Fraunhofer IAIS\\
Bonn, Germany\\
\{debattis|auer|langec\}@cs.uni-bonn.de}
}

\maketitle

\begin{abstract}
The increasing variety of Linked Data on the Web makes it challenging to determine the quality of this data, and subsequently to make this information explicit to data consumers.
Despite the availability of a number of tools and frameworks to assess Linked Data Quality, the output of such tools is not suitable for machine consumption, and thus consumers can hardly compare and rank datasets in the order of \emph{fitness for use}.
This paper describes Luzzu, a framework for Linked Data Quality Assessment.
Luzzu is based on four major components:
(1) an \emph{extensible} interface for defining new quality metrics;
(2) an \emph{interoperable}, ontology-driven back-end for representing quality metadata and quality problems that can be reused within different semantic frameworks;
(3) a \emph{scalable} stream processor for data dumps and \textsc{sparql} endpoints; and
(4) a \emph{customisable} ranking algorithm taking into account user-defined weights.
We show that Luzzu scales linearly against the number of triples in a dataset.
We also demonstrate the applicability of the Luzzu framework by evaluating and analysing a number of statistical datasets with regard to relevant metrics.
\end{abstract}


\IEEEpeerreviewmaketitle


\section{Introduction}
\label{sec:introduction}
Large amounts of data are being exchanged between organisations and published openly as Linked Data to facilitate their reuse.
Examples include the Linked Open Data cloud, as well as RDFa and Microformats data, which are increasingly being embedded in ordinary Web pages as an effect of initiatives such as schema.org.
The LOD Cloud accounts for more than 70 billion facts~\cite{LOD-Cloud}.
RDFa or Microformats are, to some extent, embedded in approx. 30\% of all Web pages\footnote{\url{http://webdatacommons.org/structureddata/\#results-2014-1}, accessed 2015-10-12}, thus particularly adding to the volume and variety dimensions of Big Data.
This emerging Web of Data shares many characteristics with the original Web of Documents.
As on the Web of Documents, we have highly varying quality on the Web of Data~\cite{journals/semweb/HitzlerJ13}.
Quality on the Web of Documents is usually measured indirectly using techniques such as the page rank.
The reason for this is that document quality is often only assessable subjectively, and thus an indirect measure such as the number of links created by others to a certain Web page is a good approximation of quality.
On the Web of Data the situation can be deemed to be both simpler and more complex at the same time. 
There is a large variety of dimensions and measures of data quality~\cite{Zaveri2012:LODQ} that can be computed automatically, so we do not have to rely on indirect indicators alone.
On the the other hand, the assessment of quality in terms of \emph{fitness for use}~\cite{Juran1974} with respect to a given use case is more challenging.

For example, data extracted from semi-structured sources, such as DBpedia, often contains inconsistencies as well as misrepresented and incomplete information.
However, in the case of DBpedia, the data quality is perfectly sufficient for enriching Web search with facts or suggestions about topics of general interest, such as entertainment topics.
For developing a medical application, on the other hand, the quality of DBpedia is probably insufficient, as shown in~\cite{zaveri2013}, since data is extracted from a semi-structured source created in a crowdsourcing effort (i.e.\ Wikipedia).
It should be noted that even the traditional document-oriented Web has content of varying quality but is still commonly perceived to be extremely useful.
Consequently, a key challenge is to determine the quality of datasets published on the Web and make this quality information explicit.
Assuring data quality in a scalable way is a challenge in Linked Data as the underlying data stems from many autonomous, evolving and increasingly large data sources.

Assessing the quality of data usually requires a large number of quality measures to be computed rather than one single measure.
Linked Data quality can be measured along several dimensions, including accessibility, interlinking, performance, syntactic validity or completeness (cf. the linked data quality survey~\cite{Zaveri2012:LODQ} for a comprehensive discussion).
In each of these dimensions, we can define a number of concrete metrics, which can be used to precisely and objectively measure a certain indicator for linked data quality.
Additionally, domain specific quality metrics can be defined, such as the number of links to an authoritative dataset.

In this article, we present a framework for assessing linked data quality, with the goal of being \emph{scalable}, \emph{extensible}, \emph{interoperable}, and \emph{customisable}.
Regarding scalability, we follow a stream processing approach, whilst for extensibility we provide an easy, declarative domain specific language for the integration of various quality measures.
With regard to interoperability, Luzzu is accompanied by a set of ontologies for capturing quality related information for reuse, including quality measures, issues and reports, that can be reused in other semantic frameworks and tools.
Even with the possibility to automatically compute quality measures, the large number of quality dimensions and measures complicates the user's task of judging whether a dataset is fit for use.
We address this problem by developing an approach for a user-driven quality-based weighted ranking of datasets, allowing users to select and give importance to some custom quality measures over others (\emph{customisable}).

The main contributions of this article are:
\begin{enumerate}
\item the open source Luzzu quality assessment framework for Linked Data (cf. Section~\ref{sec:framework}), which includes a comprehensive library of implemented quality metrics, a declarative language for creating additional domain specific quality metrics, and a set of comprehensive ontologies for capturing and exchanging data quality related information;
\item a proof of the framework's applicability to a number of statistical linked open datasets against relevant quality metrics (cf. Section~\ref{sec:ldq_eval}).
\end{enumerate}

In Section~\ref{sec:pref_eval} we evaluate the performance of the stream processors.
Section~\ref{sec:relatedwork} gives an overview of related quality assessment approaches for Linked Data, whilst in Section~\ref{sec:metricCoverage} we report the current metric coverage.
We conclude with an outlook to future work and final remarks in Section~\ref{sec:conclusion}.

\section{Quality Assessment Framework}
\label{sec:framework}

Luzzu\footnote{Sources: \url{https://github.com/EIS-Bonn/Luzzu}; Website: \url{http://eis-bonn.github.io/Luzzu/}} is a quality assessment framework for Linked Data.
The rationale of Luzzu is to provide an integrated platform that: 
(1) assesses Linked Data quality using a library of generic and user-provided domain specific quality metrics in a scalable manner; 
(2) provides queryable quality metadata on the assessed datasets; 
(3) assembles detailed quality reports on assessed datasets.
Furthermore, we aim to create an infrastructure that:
\begin{itemize}
	\item can easily be extended by users by defining custom, domain-specific metrics;
	\item implements quality-driven dataset ranking algorithms facilitating use-case driven discovery and retrieval.
\end{itemize}

\begin{figure}[t]
	\center
	\includegraphics[width=\columnwidth]{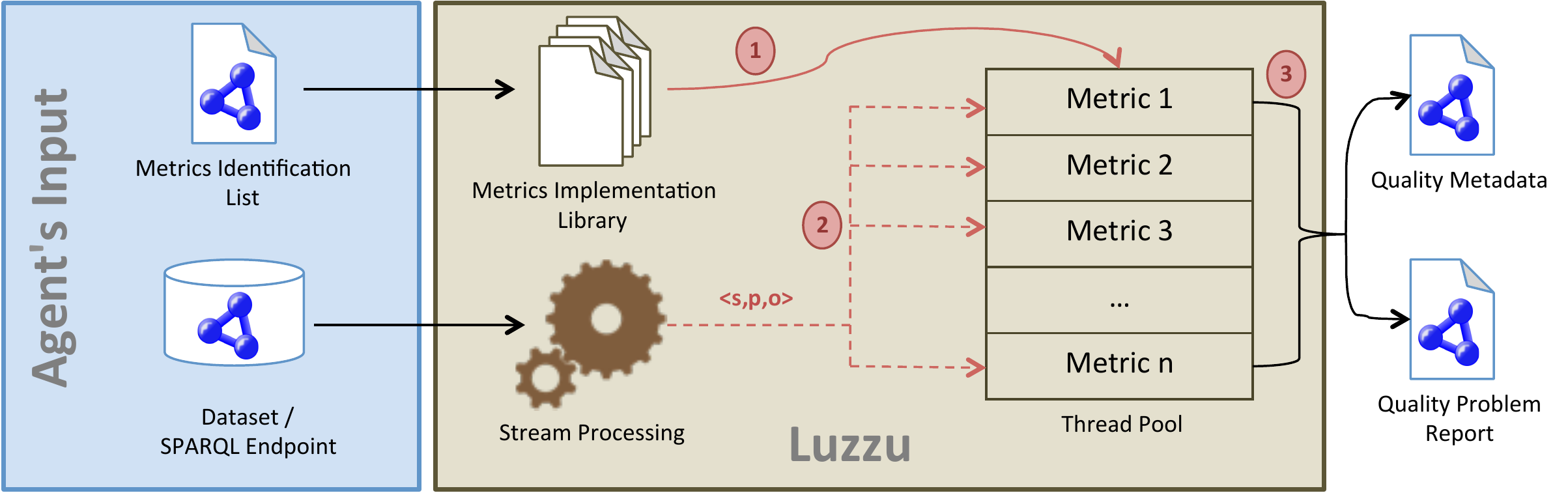} 
	\vspace{-0.5cm}
	\caption{Quality Assessment Workflow – The left side depicts the input sources from the agent, whilst the right side shows the assessment process.} 
	\label{fig:workflow}
	\vspace{-0.5cm}
\end{figure}

Figure~\ref{fig:workflow} illustrates the workflow of Luzzu.
When an agent (machine or human) initiates quality assessment for a Linked Dataset or a \textsc{sparql} endpoint, it selects a number of quality metrics.
Agents can reuse existing quality metrics or define new ones (cf. Section~\ref{sec:qualityMetricDefinition}).
The chosen metrics are then initialised, each being executed in an individual thread and initialising any required objects, such as loading a gold standard.
A processor streams triples, either from the linked dataset data dump (any RDF serialisation or compressed HDT format~\cite{AFMPG:11}) or the \textsc{sparql} endpoint, to each initialised metric (cf. Section~\ref{sec:processingUnit}).
Once all triples are processed, a quality assessment value for each configured metric is calculated.
These values are then stored as quality metadata for the assessed dataset, whilst problematic triples are reported back to the agent (cf. Section~\ref{sec:odf}).

\subsection{Quality Metric Definition}
\label{sec:qualityMetricDefinition}
A quality metric comprises an assessment function which, given a concrete RDF triple, checks if one or more constraints are satisfied and triggers one or more actions.
In Luzzu, quality metrics can be created in two ways: 
1) by implementing a Java class adhering to the specified quality metric interface, or 
2) by using the declarative \emph{Luzzu Quality Metric Language} (LQML).

LQML is a domain specific language (DSL) that enables knowledge engineers to declaratively define quality metrics whose definitions can be understood more easily.
It offers shorthand notations, abstractions and expressive power, focusing on the representation of quality metrics for linked dataset assessment. 
LQML is designed in a way that metrics can be written by non-programmers that are experts in the domain  \cite{vanDeursen:2000:DLA:352029.352035,dsl-little-languages}. 
Hudak~\cite{dsl-little-languages} suggests that DSLs have the potential to improve productivity in the long run and with LQML we aim to simplify Linked Data quality assessment.
In order not to limit LQML expressiveness, programmers can extend it by custom functions for complex tasks, such as network operations (e.g. \texttt{isDereferenceable}) or logical consistency checks (e.g. \texttt{hasValidInverseFunctionalPropertyUsage}), which can then be used in a metric definition's conditions and actions.
This follows the best-practice of other domain specific languages, such as \textsc{sparql} or \textsc{XPath}, which support the use of user-defined functions.
Listing~\ref{lst:lqmlExample} shows an LQML defined metric that checks if the subject and the object of a triple are dereferenceable (i.e. has a \texttt{303 See Other} HTTP code, or is a hash URI).

\begin{lqml}[Using Custom Functions in LQML]{lst:lqmlExample}
def{Dereferenceability}:
  metric{<http://purl.org/eis/vocab/dqm#Dereferenceablity>};
  label{"Dereferenceability of Resources"};
  description{"Measures the @ratio@ of valid dereferenceable resources"};
  x = match{(isURI(?s) && isDereferenceable(?s))}
    => action{count(unique(?s))};
  y = match{(isURI(?o) && isDereferenceable(?o))}
    => action{count(unique(?o))};
  finally{ratio(add(action(x), action(y)), totaltriples(?s)}.
\end{lqml}

Although Luzzu provides a \textsc{sparql} endpoint processor, no metric is implemented as a \textsc{sparql} query.
The reason behind opting against \textsc{sparql} metrics (for example as used in \cite{kontokostasDatabugger}) is that constraints checked by metrics implemented in \textsc{sparql} are limited to the data and its schema, thus ignoring other important linked data quality measures, such as the identification of outliers or detecting resource performance issues.
Nevertheless, any metric that can be expressed in a \textsc{sparql} query can be defined as a metric for Luzzu.

\subsection{Processing Linked Datasets}
\label{sec:processingUnit}
Luzzu provides a scalable RDF processor which streams a dataset's triples into all initialised metric processors (cf. Figure~\ref{fig:workflow} (label 2)).
Streaming ensures scalability beyond the limits of main memory, and \emph{parallelisability}, since the parsing of a dataset can be split into several streams to be processed on different threads, cores or machines.

The input dataset can be either a serialised linked data dump or a \textsc{sparql} endpoint.
Any such RDF data source is processed triple by triple.
In addition, the framework includes a \emph{Spark}\footnote{\url{http://spark.apache.org/}} processor (an in-memory equivalent to Hadoop), thus enabling Luzzu to exploit Big Data infrastructures to their full potential.
The rationale is to \emph{map} the processing of large datasets on multiple clusters, and then using a \emph{reduce} function, populating a queue that feeds the metrics.
Metric computations are not “associative” in general, which is one of the main requirements to implement a MapReduce job; therefore, instantiated metrics are split into different threads in the master node.

\paragraph{Processing SPARQL Endpoints}
\label{sec:ProcSPARQLEndpoints}
Luzzu provides a processor to assess the quality of a linked data knowledge base (KB) from a \textsc{sparql} endpoint when dumps are not available.
For this job, triple statements are fetched from the KB and streamed to the metrics in the thread pool.
Public \textsc{sparql} endpoints often truncate results for queries that are very expensive to compute.
For example, a query asking for the entire KB in the public DBpedia endpoint will return only the first 10,000 results.
Therefore, statements are retrieved in intervals using the \textsc{sparql} \texttt{OFFSET} and \texttt{ORDER BY} modifiers.
Assessing the data quality of a knowledge base from a \textsc{sparql} endpoint comes with further challenges.
For example, if a \textsc{sparql} \emph{write} transaction (insert or update) is performed during the assessment, the whole quality assessment might be compromised.
Thus, we discourage the use of \textsc{sparql} endpoints where possible.

\subsection{Ontology-Driven Framework}
\label{sec:odf}
\begin{figure}[tb]
\center
\includegraphics[width=\columnwidth]{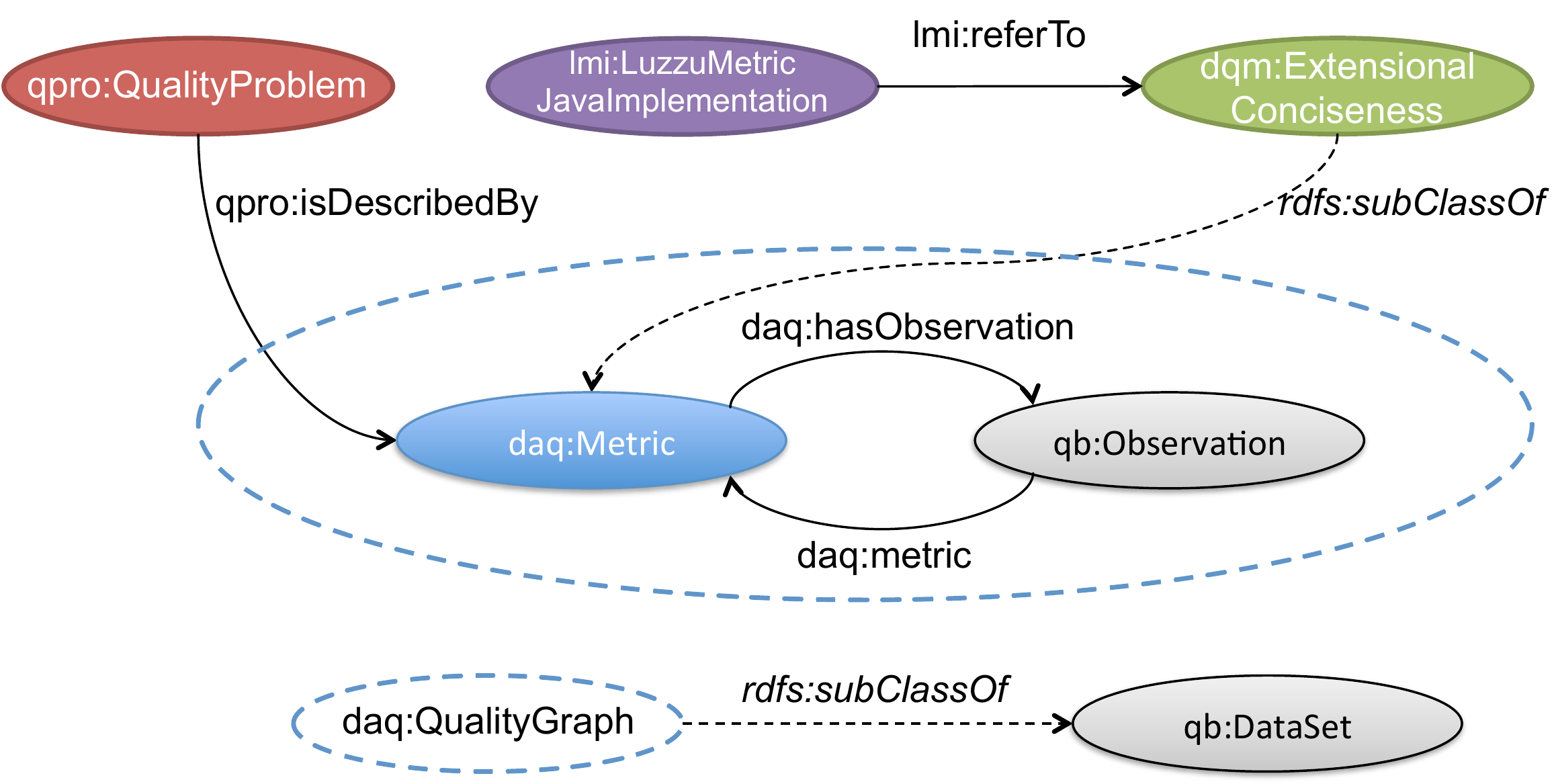} 
\vspace{-0.5cm}
\caption{Key relationships in the data quality ontologies} 
\label{fig:ontology_relationships}
\end{figure}

Luzzu~employs an underlying semantic knowledge layer to capture the quality assessment results.
The \emph{semantic schema layer} consists of a \emph{representational} (split into generic and specific sub-levels) and an \emph{operational} level.
\begin{figure}[tb]
\center
\includegraphics[width=\columnwidth]{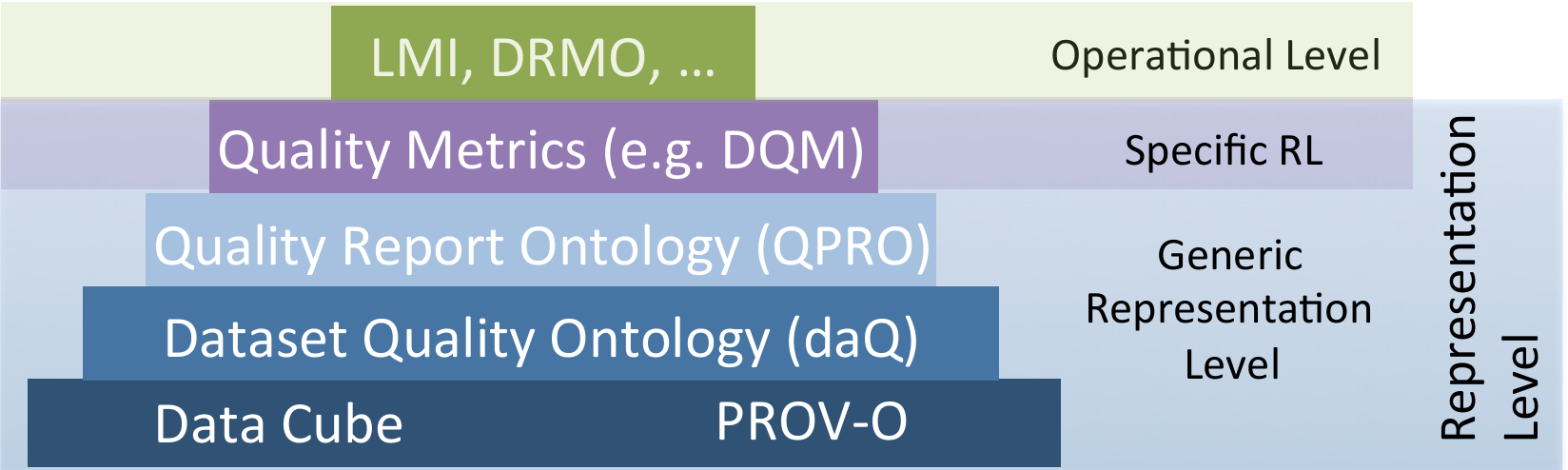} 
\vspace{-0.5cm}
\caption{Ontology Stack} 
\label{fig:ontology_framework}
\end{figure}
Figure~\ref{fig:ontology_framework} shows the framework's ontology schema stack, where the lower level comprises generic ontologies that form the foundations of the quality assessment framework, and the upper level comprises specific ontologies required for the various quality assessment tasks.
The ontologies used in this stack allow for a comprehensive and holistic representation of linked data quality information.
The daQ core ontology is based on the W3C RDF Data Cube~\cite{w3c:vocab-data-cube} and PROV-O~\cite{w3c:prov-o} vocabularies.
Figure~\ref{fig:ontology_relationships} depicts the relationships between the ontologies.\\

\subsubsection{Generic Representation Level}
The generic representation level is domain independent, and can be easily reused in similar frameworks for assessing data quality.
The two vocabularies in this level are the Dataset Quality Ontology providing queryable metadata and the Quality Problem Report Ontology for assembling detailed quality reports (cf. Figure~\ref{fig:workflow} label 3).

The aim of the \emph{Dataset Quality Ontology (daQ)}~\cite{DebattistaEtAl:daQDataCube2014}\footnote{All Luzzu ontologies use the namespace \url{http://purl.org/eis/vocab/{prefix}}. 
For each ontology then one should add the relevant ontology prefix, e.g., \emph{daq}.} is to provide a core vocabulary that defines how quality metadata should be represented at an abstract level.
It is used to attach the results of quality assessment of a Linked Dataset to the dataset itself. 
daQ is the core vocabulary of this schema layer, and any ontology describing quality metrics added to the framework (in the specific representation level) should extend it.
The benefit of having an extensible schema is that quality metrics can be added to the framework without major changes, as the representation of new metrics would follow those defined previously.

The \emph{Quality Problem Report Ontology (QPRO)} enables the fine-grained description of quality problems found while assessing a dataset.
It comprises two core classes: \texttt{qpro:QualityReport}, representing a report on the problems detected during the assessment of quality on a dataset, and \texttt{qpro:QualityProblem}, representing the individual quality problems contained in that report. 
Each \texttt{qpro:QualityProblem} is described by the following properties:
\begin{itemize}
	\item \texttt{qpro:computedOn} refers to the URI of the dataset on which a certain quality assessment has been performed. 
		This property is attached to a \texttt{qpro:QualityReport}. 
	\item \texttt{qpro:hasProblem} identifies problem instances in a report and links a \texttt{qpro:QualityProblem} to a \texttt{qpro:QualityReport}. 
	\item \texttt{qpro:isDescribedBy} describes each \texttt{qpro:QualityProblem} using an instance of a \texttt{daq:Metric}.
	\item \texttt{qpro:problematicThing} represents the actual problematic instance from the dataset. 
		This can be a list (\texttt{rdf:Seq}) of resources or of reified RDF statements.
	\item \texttt{qpro:inGraph} refers to the assessed graph, since quality assessments can be performed on multiple graphs.
\end{itemize}
The problem report is envisaged to be used by linked data cleaning tools, as it identifies all problematic triples.

\subsubsection{Specific Representation Level and Operational Level}
The specific representation level consists of semantically defined quality metrics, based on the three layers of the abstract level of daQ.
Luzzu provides a small configuration vocabulary, the Luzzu Metric Implementation (LMI) vocabulary, which enables the linking between a semantically defined metric and its Java or LQML implementation.
The semantic definition, together with its concrete implementation, is part of the framework's library, which facilitates reuse.
Complementing Luzzu, we have implemented more than 40 linked data quality metrics\footnote{These can be downloaded from \url{https://github.com/diachron/quality}}, most of which were identified in~\cite{Zaveri2012:LODQ}.\\

\subsection{User-Driven Ranking}
\label{sec:framework_operations-unit}
In the spirit of \emph{fitness for use}, we propose a \emph{user-driven ranking algorithm} that enable users to define weights on their preferred quality categories, dimensions or metrics, as deemed suitable for the task at hand.
User-driven ranking is formalised as follows:

Let $v: F_m\to \{ \mathbb{R} \cup \mathbb{N} \cup \mathbb{B} \cup \ldots \}$ be the function that yields the value of a metric – most commonly a real number, but it could also be an integer, a boolean, or any other simple type.

\paragraph{Ranking by Metric}
Ranking datasets by individual metrics requires computing a weighted sum of the values of all metrics chosen by the user.
Let $m_i$ be a metric, $v(m_i)$ its value, and $\theta_i$ its weight ($i=1,\dots,n$), then the weighted value $v(m_i, \theta_i)$ is given by:

\begin{Def1}[Weighted metric value]
\label{def:metric}
\begin{align*}
  v(m_i, \theta_i) := \theta_i \cdot v(m_i)
\end{align*}
\end{Def1}

The \emph{sum} $\sum_{i=1}^n v(m_i, \theta_i)$ of these weighted values, with the same weights applied to all datasets under assessment, determines the ranking of the datasets.

\paragraph{Ranking by Dimension}
When users want to rank datasets in a less fine-grained manner, they assign weights to quality dimensions.\footnote{Zaveri et al.\ define, for example, the dimension of licensing, which comprises the metrics ``[existence of a] machine-readable indication of a license'' and ``human-readable indication of a license''~\cite{Zaveri2012:LODQ}.}
The weighted value of the dimension $D$ is computed by evenly applying the weight $\theta$ to each metric $m$ in the dimension.

\begin{Def1}[Weighted dimension value]
\label{def:dimension}
\begin{align*}
v(D, \theta) := \frac{\sum_{m\in D} v(m, \theta)}{\#D} = \theta \frac{\sum_{m\in D} v(m)}{\#D}
\end{align*}
\end{Def1}

\paragraph{Ranking by Category}
A category $C$ is defined to comprise one or more dimensions $D$\footnote{In the classification of Zaveri et al., ``licensing'' is a dimension within the category of ``accessibility dimensions''~\cite{Zaveri2012:LODQ}.}, thus, similarly to the previous case, ranking on the level of categories requires distributing the weight chosen for a category over the dimensions in this category and then applying Definition~\ref{def:dimension}.

\begin{Def1}[Weighted category value]
\label{def:category}
\begin{align*}
  v(C, \theta) := \frac{\sum_{D\in C} v(D, \theta)}{\#C}
\end{align*}
\end{Def1}



\section{Performance Evaluation}
\label{sec:pref_eval}
The aim of this experiment is to assess the scalability of the Luzzu framework. 
Runtime is measured for both processors against a number of datasets ranging from 10k to 125M triples.
Since the main goal of the experiment is to measure the processors' performance, having datasets with different quality problems is considered to be irrelevant at this stage.
Therefore we generated synthetic datasets of different sizes using the Berlin SPARQL Benchmark (BSBM) V3.1 data generator\footnote{BSBM is primarily used as a benchmark to measure the performance of SPARQL queries against large datasets; cf.\ \url{http://wifo5-03.informatik.uni-mannheim.de/bizer/berlinsparqlbenchmark/spec/}.}.
We generated datasets with a scale factor of 24, 56, 128, 199, 256, 666, 1369, 2089, 2785, 28453, 70812, 284826, 357431, which translates into approximately 10K, 25K, 50K, 75K, 100K, 250K, 500K, 750K, 1M, 10M, 25M, 50M, 100M, and 125M triples respectively.
The stream and SPARQL processor tests were performed on a Unix based machine with an Intel Core i5 2.4GHz and 4GB of RAM, whilst for the Spark processor three worker clusters were set up.

\begin{figure}[tb]
\center
\includegraphics[width=\columnwidth]{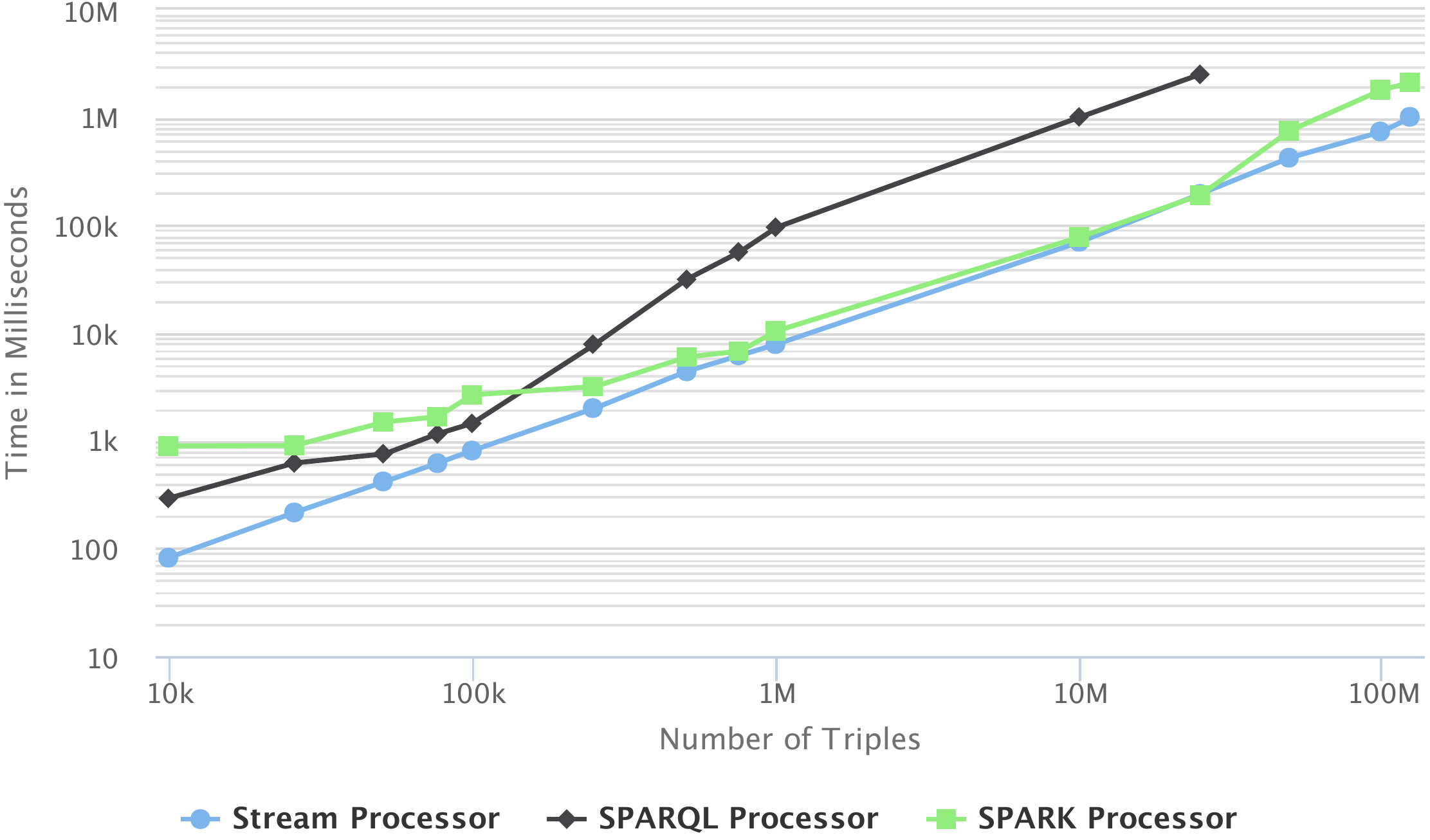} 
\vspace{-.5cm}
\caption{Time vs. dataset size in triples} 
\label{fig:performance}
\end{figure}

\paragraph{Results}
Figure~\ref{fig:performance} shows the time taken (in ms) to process datasets of different sizes.
We normalised the values with a log (base 10) function on both axes to improve readability.
All processors scale linearly as the number of triples grows.
The SPARQL processor was not responding in acceptable time for datasets larger than 25M triples.
The results also confirm the assumption that big data technologies such as Spark are not beneficial for smaller datasets, whilst processing data from a SPARQL endpoint takes more time than the other two processing approaches.

From these results, we also conclude that for up to 125M triples, the stream processor performs better than the Spark processor.
A cause for this difference is that the Spark processor has to deal with the extra overhead needed to enqueue and dequeue triples on an external queue, however, as the number of triples increases, the performance of both processors converges.
With an increasing number of metrics to be computed, the execution time of both processors increases but remains linear w.r.t.\ the size of the dataset.

\section{Statistical Linked Dataset Quality Analysis}
\label{sec:ldq_eval}

To showcase Luzzu in a real-world scenario, a Linked Data quality evaluation was performed\footnote{Assessment done on dataset dumps as at July 2015.} on nine statistical linked datasets from 270a.info.
These datasets were assessed in Luzzu over a number of quality metrics.
Results for these datasets can be found online\footnote{\url{http://tinyurl.com/icsc2016-luzzu}} and on Luzzu Web\footnote{\url{http://purl.org/net/Luzzu}}, where metadata can be visualised and datasets can be ranked and filtered.

\subsection{Chosen Metrics}
\label{sec:ldq_eval_chosen_metrics}
25 metics from ten different dimensions were used for this assessment.
Whilst most of these metrics were identified from~\cite{Zaveri2012:LODQ}, two provenance related metrics were additionally defined for this assessment.
The \emph{W3C Data on the Web Best Practices working group}\footnote{\url{https://www.w3.org/2013/dwbp/} (the first author is a member)} has defined a number of best practices in order to support the self-sustained eco-system of the Data Web.
With regard to provenance, the working group argues that provenance information should be accessible to data consumers, whilst they should also be able to track the origin or history of the published data~\cite{W3C:DWBP:BestPractices:WorkingDraft}.
In light of this statement, we defined the \emph{Basic Provenance} and \emph{Extended Provenance} metrics.
The \emph{Basic Provenance} metric checks if a dataset, usually of type \texttt{void:Dataset} or \texttt{dcat:Dataset}, has the most basic provenance information; that is information related to the creator or publisher, using the \texttt{dc:creator} or \texttt{dc:publisher} properties.
On the other hand, the \emph{Extended Provenance} metric checks if a dataset has the required information that enables a consumer to track down the origin.
Each dataset should have an entity containing the identification of a \texttt{prov:Agent} and the identification of one or more \texttt{prov:Activity}.
The identified activities should also have an agent associated with it, apart from having a data source attached to it.

\begin{figure}[tb]
\center
\includegraphics[width=\columnwidth]{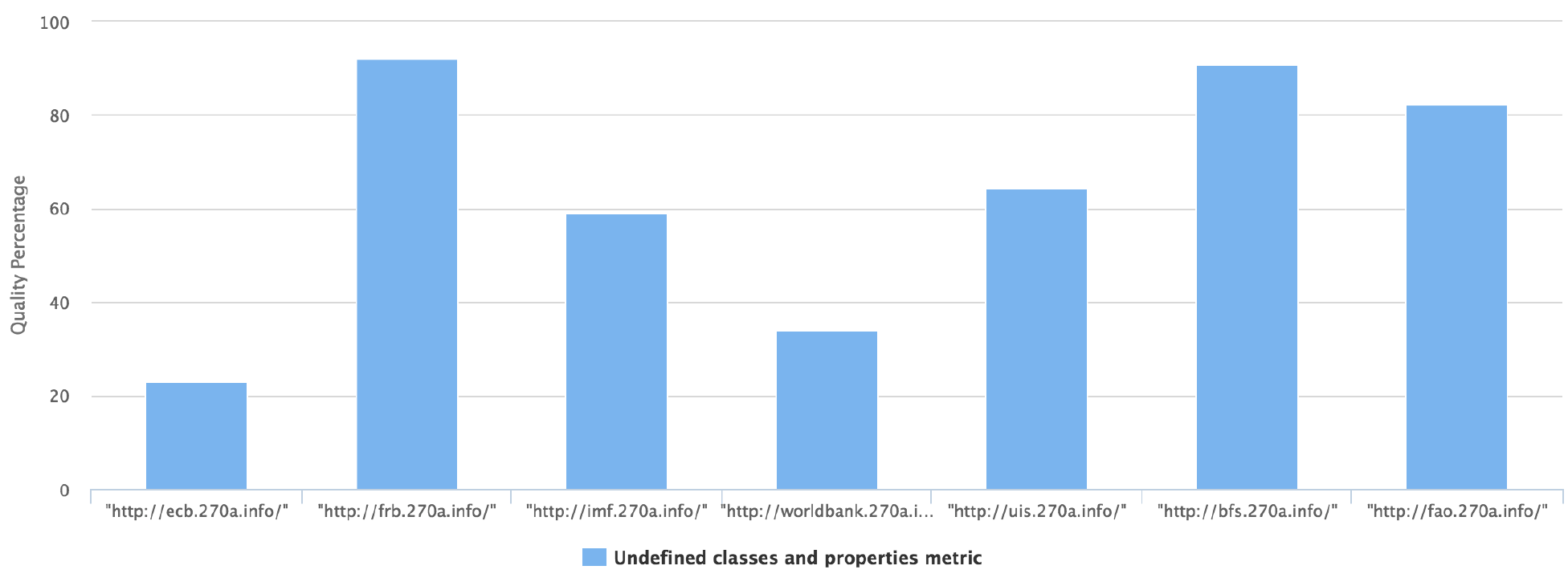} 
\caption{Quality Assessment Values for the Undefined Classes and Properties metric – 100\% means high quality.} 
\vspace{-.5cm}
\label{fig:undef}
\end{figure}

Figure~\ref{fig:undef} depicts the quality assessment values for one of the assessed metrics – Undefined Classes and Properties.
The metrics were implemented as described by their respective authors (as identified in~\cite{Zaveri2012:LODQ}), expect for the \emph{Misreported Content Types}, and \emph{No Entities as Members of Disjoint Classes} metrics, where approximation techniques, more specifically reservoir sampling, were used to improve scalability~\cite{DBLP:conf/esws/DebattistaL0A15}.

\subsection{Results Discussion}
In general, the datasets adhere to high quality when it comes to the \emph{consistency}, \emph{representational conciseness}, and \emph{performance}  dimensions. 
The consistency dimension relates to the efficiency of the system hosting the dataset, whilst the representational conciseness dimension deals with data clarity and compactness.
In other words, the latter dimension measures if
(1) resource URIs are short, that is a URI should be maximum 80 characters\footnote{This best practice is mentioned at \url{http://www.w3.org/TR/chips/\#uri}} and have no parameters; and 
(2) a dataset is free of RDF Collections and Lists.

The performance metrics measure the extent to which the dataset is free of logical and formal errors and contradictions.
According to the \emph{Cool URIs guidelines}\footnote{\url{http://www.w3.org/TR/cooluris/}}, Hash URIs should be used in small and stable datasets since they provide stability with regard to the retrieval performance, since no extra 303 redirects are required.
On the other hand, Slash URIs should be used in large datasets.
Following~\cite{Flemming2011}, the minimum number of triples required to be considered as a “large dataset” is set to 500K.
The assessed transparency data space has around 43K triples, of which only 30 URIs are hash URIs.

Whilst all data spaces provided basic provenance information, tracking the origin of the data source (an extended provenance metric) is not possible in all datasets.
For example, in the Transparency data space, no provenance information is recorded.
Furthermore, although provenance was recorded in the other assessed data spaces, entities might have had any of the required attributes mentioned in the previous section missing.
This was confirmed by a SPARQL query on the ECB dataset, where from 460 entities only 248 were attributable to an agent, whilst each activity contained pointers to the data source used and the agent it is associated with. 

With regard to the \emph{timeliness} metrics, it was noticed that the datasets were missing the information required for the assessment.
For example, the \emph{timeliness of resource} metric required the last modified time of the original data source and the last modified time of the semantic resource.
Similarly, for the \emph{freshness of dataset} and \emph{currency of dataset} (recency) metrics, the datasets had missing meta information, except for the Transparency and World Bank datasets for the latter mentioned metric.
These still gave a relatively a low quality score, and a manual investigation revealed that the datasets had not been updated for over a year.

In order to measure the \emph{interoperability} metrics, some pre-processing was required.
All nine datasets were tagged with a number of domains (such as Statistical, Government and Financial) and a number of vocabularies that were deemed suitable.
For each domain, the \emph{Linked Open Vocabulary API} (LOV\footnote{\url{http://lov.okfn.org/dataset/lov/}}) was queried, and resolvable vocabularies were downloaded to Luzzu.
Although one does not expect that datasets reuse all existing and available vocabularies, the reuse of vocabularies is encouraged.
In order to understand the problem better, a SPARQL query, getting all vocabularies referred to with the \texttt{void:vocabulary} property, was executed against one of the data spaces (more specifically the BFS) endpoint.
It resulted in nine vocabularies, five of which were removed since they were the common OWL, RDF, RDFS, SKOS and DCMI Terms vocabularies.
The rest of the vocabularies were \url{http://bfs.270a.info/property/}, \url{http://purl.org/linked-data/sdmx/2009/concept}, \url{http://purl.org/linked-data/xkos}, \url{http://www.w3.org/ns/prov}.
The LOV API suggested 14 vocabularies when given the terms assigned to this dataset, none of which appeared in the list of vocabularies used.
One possible problem of this metric is that since these are qualitative metrics, which measure quality according to a perspective, the dataset might have been tagged wrongly.
On the other hand, the metric relies on an external service, LOV, which might 
(1) not have all vocabularies in the system, 
(2) vocabularies might not appear in the results since the search term is not included in the vocabulary description or label.

With all data dumps and SPARQL endpoint URIs present in the VoID file of all datasets, the 270a.info data space publishers are well aware of the importance of having the data available and obtainable (\emph{availability dimension}).
The datasets also fare well with regard to having a relatively low number of misreported content types for the resource URIs they use.
The percentage loss in this metric is attributable to the fact that in this metric all URI resources (local, i.e.\ having a base URI of 270a.info, and external) that return a \texttt{200 OK} HTTP status code are being assessed. 
In fact, most of the problematic URIs found were attributable to the usage of the following three vocabularies:
\url{http://purl.org/linked-data/xkos};
\url{http://purl.org/linked-data/cube};
\url{http://purl.org/linked-data/sdmx/2009/dimension}.
Manual inspection using the \texttt{curl} command-line HTTP client shows that the content type returned by these is \texttt{text/html; charset=iso-8859-1}, even when enforcing a content negotiation for semantic content types such as \texttt{application/rdf+xml}.
When assessing the World Bank dataset, there was also one instance (more specifically \url{http://worldbank.270a.info/property/implementing-agency}) that the metric incorrectly reported as a problematic resource URI. 
It was observed that the mentioned resource took an unusual long time to resolve, which might have resulted in a timeout. 

All data spaces have a machine readable license (\emph{licensing dimension}) in their VoID file, more specifically those datasets (\texttt{void:Dataset}) that contain observation data.
The datasets that describe metadata, such as \url{http://bfs.270a.info/dataset/meta}, have no machine readable license attributed to them, thus such cases decreased the quality result for this metric.
On the other hand, the datasets had no human readable licenses in their description. 
This metric verifies whether a human-readable text, stating the licensing model attributed to the resource, has been provided as part of the dataset.
In contrast to the machine readable metric, literal objects (attached to properties such as \texttt{rdfs:label}, \texttt{dcterms:description}, \texttt{rdfs:comment}) are analysed for terms related to licensing.

The \emph{versatility dimension} refers to the different representation of data for both human and machine consumption.
With regard to human consumption (multiple language usage), the BFS dataset on average supports four languages, the highest from all assessed datasets.
This metric checks all literals that have a language tag.
On the other hand, data spaces offer different serialisations (attributed via the \texttt{void:feature} -- two formats for each data space), but only one dataset in each data space (e.g. \url{http://bfs.270a.info/dataset/bfs}) has this property as part of its VoID description.
If, for example, we take the OECD data space, there are 140 defined VoID datasets, but only one of them has the \texttt{void:feature} property defined.

The \emph{interpretability dimension} refers to the notation of data, that is, checking if a machine can process data.
Therefore, a dataset should have a low blank node usage, since blank nodes cannot be resolved and referenced externally.
In this regard, all datasets have a low usage, and these are generally used to describe \texttt{void:classPartition} in the VoID metadata.
The datasets also use classes and properties that are defined in ontologies and vocabularies (cf. Figure~\ref{fig:undef}), with an occasional usage of the typical \url{http://example.org}, such as \url{http://example.org/XStats\#value}.


\section{Related Work}
\label{sec:relatedwork}

\subsection{Quality Assessment Frameworks}
Table~\ref{tbl:comparison} gives an overview on the tools discussed in this section.
As can be seen in this table Luzzu, with its focus on scalability, extensibility and quality metadata representation, fills a gap in the space of related work.

\begin{table}[tb]
\tiny
\setlength\tabcolsep{.1cm}
\resizebox{\columnwidth}{!}{%
    \begin{tabular}{l|c|c|c|c|c|c|c}
    ~                                     &     \emph{Flemming}              & \emph{LinkQA}                   & \emph{Sieve}                & \emph{RDF Unit}                & \emph{Triple Check Mate}             & \emph{LiQuate}                                & \emph{Luzzu}                                          \\ \hline
    \emph{Scalability}        & \ding{55}                        & \ding{51}                      & \ding{51}                  & \ding{51}                     & Crowdsourcing & N/A & \ding{51}                                            \\\hline
    \emph{Extensibility} & \ding{55}                        & Java       & XML & SPARQL & No                            & Bayesian rules & Java, LQML \\\hline
    \emph{Quality Metadata}              &\ding{55}                        &\ding{55}                       & \ding{51} (Optional)       & \ding{51}                     & \ding{55}                            & \ding{55}                                     & \ding{51}                                            \\\hline
    \emph{Quality Report}                & HTML & HTML & \ding{55}                   & HTML, RDF        & \ding{55}                            & \ding{55}                                     & RDF                                      \\\hline
    \emph{Collaboration}                         & \ding{55}                        & \ding{55}                       & \ding{55}                   & \ding{55}                      & \ding{51}                           & \ding{55}                                     & \ding{55}                                             \\\hline
    \emph{Cleaning support}                & \ding{55}                        & \ding{55}                       & \ding{51}                  & \ding{55}                      & \ding{55}                            & \ding{55}                                     & \ding{55}                                             \\\hline
    \emph{Last release}                       & 2010                      & 2011                     & 2014                 & 2014                    & 2013                          & 2014                                   & 2015                                           \\\hline
    \end{tabular}
    }
    \label{tbl:comparison}
    \caption{Functional comparison of Linked Data quality tools.}
    \vspace{-0.5cm}
\end{table}

Flemming~\cite{Flemming2011} provides a simple web user interface and a walk through guide that helps a user to assess data quality of a resource using a set of defined metrics.
Metric weights can be customised either by user assignment, though no new metric can be added to the system.
Luzzu enables users to create custom metrics, including complex quality metrics that can be customised further by configuring them, such as providing a list of trustworthy data providers.
In contrast to Luzzu, Flemming's tool outputs the result and problems as unstructured text.

\emph{LinkQA}~\cite{Gueret2012} is an assessment tool to measure the quality (and changes in quality) of a dataset using network analysis measures.
The authors provide five network measures, namely degree, clustering coefficient, centrality, sameAs chains, and descriptive richness through sameAs.
Similarly to Luzzu, new metrics can be integrated into LinkQA, but these metrics are related to topological measures.
LinkQA reports the assessment findings in HTML.
Although structured, such reports cannot be attached to LOD datasets for further machine reuse, such as running queries over the results of an earlier assessment.

\emph{Sieve}~\cite{Mendes2012} uses metadata about named graphs to assess data quality, where assessment metrics are declaratively defined by users through an XML configuration.
In such configurations, scoring functions (that can also be extended or customised) can be applied on one or an aggregate of metrics.
Whereas Luzzu provides a quality problem report that can be potentially be imported in third party cleaning software, Sieve provides its own data cleaning process, where data is cleaned based on a user configuration.
The quality assessment tool is part of the LDIF Linked Data Integration Framework, which supports Hadoop.

Similarly to Flemming's tool, the \emph{LiQuate}~\cite{Ruckhaus2014} tool provides a guided walkthrough to view pre-computed datasets.
LiQuate is a quality assessment tool based on Bayesian Networks, which analyse the quality of datasets in the LOD cloud whilst identifying potential quality problems and ambiguities.
This probabilistic model is used in LiQuate to explore the assessed datasets for completeness, redundancies and inconsistencies.
Data experts are required to identify rules for the Bayesian Network.

\emph{Triple check mate}~\cite{zaveri2013} is mainly a crowdsourcing tool for quality assessment, supported with a semi-automatic verification of quality metrics.
With the crowdsourcing approach, certain quality problems (such as semantic errors) might be detected easily by human evaluators rather than by computational algorithms.
On the other hand, the proposed semi-automated approach makes use of reasoners and machine learning to learn characteristics of a knowledge base.

\hyphenation{RDF-Unit}
\emph{RDFUnit}~\cite{kontokostasDatabugger} provides test-driven quality assessment for Linked Data.
In RDFUnit, users define quality test patterns based on a SPARQL query template.
Similar to Luzzu and Sieve, this gives the user the opportunity to adapt the quality framework to their needs.
The focus of RDFUnit is more to check for integrity constraints expressed as SPARQL patterns.
Thus, users have to understand different SPARQL patterns that represent these constraints.
Quality is assessed by executing the custom SPARQL queries against dataset endpoints.
In contrast, Luzzu does not rely on SPARQL querying to assess a dataset, and therefore can compute more complex processes (such as checking for dereferenceability of resources) on dataset triples themselves.
Test case results, both quality values and quality problems, from an RDFUnit execution, are stored and represented as Linked Data and visualised as HTML.
However, Luzzu's daQ ontology enables a more fine-grained and detailed quality metric representation.
For example, representing quality metadata with daQ enables the representation of a metric value change over time.



\section{Metric Coverage}
\label{sec:metricCoverage}

\begin{figure}[tb]
\center
\includegraphics[width=\columnwidth]{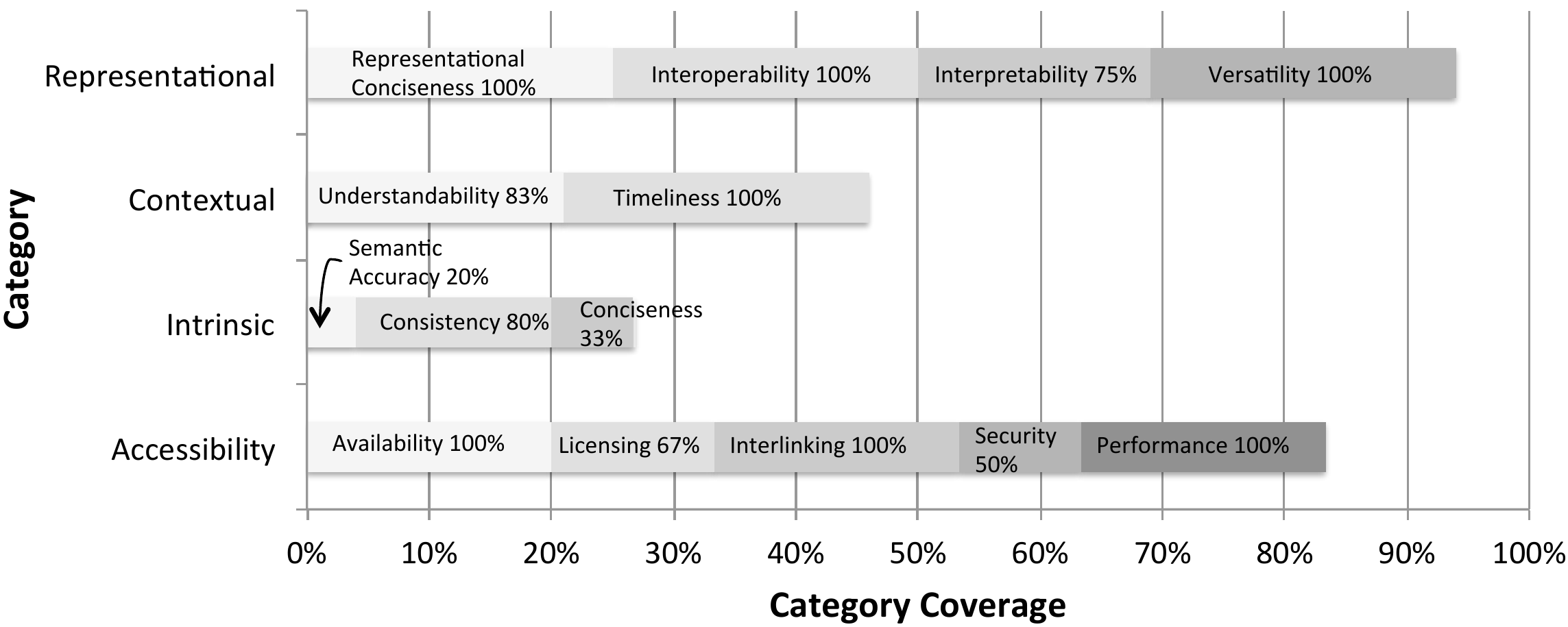} 
\vspace{-.5cm}
\caption{Dimension-Metric Coverage per Category} 
\label{fig:coverage}
\vspace{-.5cm}
\end{figure}

As part of Luzzu, we implemented a number of Linked Data quality metrics, most of which were identified in~\cite{zaveri2013}.
Some of the metrics were implemented as described by the relevant related articles, but we also implemented a number of metrics using various probabilistic and approximation techniques~\cite{DBLP:conf/esws/DebattistaL0A15}.
In total, we implemented around 60\% of the metrics, including a number of subjective metrics from~\cite{zaveri2013}.
Figure~\ref{fig:coverage} shows the complete coverage of metrics per category.
We also implemented two metrics concerning provenance, which we described in Section~\ref{sec:ldq_eval_chosen_metrics}.

Time-efficient computation of data quality metrics is aligned with the scalability goal of Luzzu.
Most of the implemented metrics are \emph{deterministic} and computable within \emph{polynomial time}.
On the other hand, once these metrics are exposed to large datasets, the metrics' upper bound grows and, as a result, the computational time becomes intractable.
Therefore, probabilistic techniques is the solution for approximating the quality value in a low running time yet giving a near-to-accurate result~\cite{DBLP:conf/esws/DebattistaL0A15}.
We used reservoir sampling to implement various metrics that require network access for its assessment.
These include \textit{Dereferenceability}, \textit{Links to External Data Providers}, \textit{Misreported Content Type}, and \textit{Member of Disjoint Classes} metrics.
Bloom filters are used to detect the number of unique instances for the \textit{Extensional Conciseness} metric.
Finally, clustering coefficient estimation is use as part of the \textit{Good Interlinks} metric.

\section{Final Remarks}
\label{sec:conclusion}

Assessing the quality of linked datasets has a social impact on the Web of Data and its users, producers as well as consumers.
Having good quality datasets ensures their reusability and thus helps in decreasing the number of duplicate and redundant resources on the Web.
The components in the Luzzu framework  impacts the Web of Data positively with regard to data quality assessment.
From a research perspective, Luzzu provides a framework for which data quality researchers can implement different metrics without the need of developing their own framework for processing.

We have developed Luzzu, an \emph{extensible}, \emph{scalable}, \emph{interoperable} and \emph{customisable} framework as a means to assess data quality in Linked Datasets.
This framework encompasses four major components:
(1) an interface for adding custom quality metrics in Luzzu, using either traditional Java classes or the LQML domain specific language;
(2) lightweight vocabularies to represent quality metadata and quality problems; 
(3) a big-data ready dataset and \textsc{sparql} endpoint processor; and 
(4) a user-driven quality-based weighted ranking algorithm.

The main contribution of the first component is that with the Luzzu framework we ensured that datasets in various domains with different schemas can be assessed, due to its nature of being extensible.
Furthermore, with the introduction of LQML, we aim to simplify Linked Data quality assessment in the long run.
The vocabularies (second component) developed for Luzzu are novel with regard to data quality on the Web of Data, whilst enhancing interoperability across different quality frameworks.
Whilst the quality report vocabulary aims at bridging the gap between data quality assessment and cleaning/repairing tools, the contributions of the Data Quality Ontology (daQ) are already fruitful as the W3C Data on the Web Best Practices WG is considering to adopt this lightweight vocabulary as a core module of their Data Quality Vocabulary~\cite{w3c:dwbp:dqv}.
The streaming approach (third component) of the Luzzu processors ensures that large datasets can be fully assessed in a linear fashion, whilst the introduction of the Spark processor enables the realisation of the full potential of Big Data infrastructures.
Finally, our contribution for the last component is mainly targeted towards data consumers, who can now filter and rank quality assessed datasets, to find a `fit' dataset, based on their specific quality priorities.
The Luzzu Web portal makes this ranking procedure usable in a straightforward way.
In order to make Luzzu available immediately, we implemented a number of quality metrics that can be used to assess linked datasets.

Overall, the key contribution of this research stems from the integration of these four components to create a \textbf{holistic} framework that enables the whole process of quality assessment in Linked Data with
(1) the aim of helping data consumers to separate the wheat from the chaff to find \emph{fit for use} datasets; and
(2) the possibility to include the framework in a stack of tools for co-evolution and curation of linked datasets.

We see Luzzu as the first step on a long-term research agenda aiming at shedding light on the quality of data published on the Web. 
In this paper we provided a thorough discussion regarding the quality of a number of statistical Linked Datasets, as a first step towards this goal.
In future, we also plan to investigate techniques for incremental quality assessment, avoiding the reassessment of the whole dataset after curation.


\section{Acknowledgment}
This work is supported by the European Commission under the Seventh Framework Program FP7 grant 601043 (\url{http://diachron-fp7.eu}).

\printbibliography

\end{document}